\documentclass{appolb}
\usepackage{epsfig}
\begin{document}
\title{Constraining the Selectron Mass  in the Process  $e^- + \gamma 
\longrightarrow \tilde{\chi}_1^0 +\tilde{e}_{L/R}^- \longrightarrow e^- 
\tilde{\chi}_1^0 \tilde{\chi}_1^0$
\thanks{Presented at XIII. Summer School of Theoretical Physics, Ustron}%
}
\author{Claus Bl\"ochinger\footnote{email:bloechi@physik.uni-wuerzburg.de}
\and
Hans Fraas\footnote{email:fraas@physik.uni-wuerzburg.de}
\address{Institut f\"ur Theoretische Physik, Universit\"at W\"urzburg, Am 
Hubland, \\
D-97074 W\"urzburg, Germany}
}
\maketitle

\begin{abstract}
With the process $e^-\gamma \rightarrow \tilde{\chi}_1^0 \tilde{e}_{L/R}^- \rightarrow e^- \tilde{\chi}_1^0 \tilde{\chi}_1^0$ it is possible to constrain selectron masses above the kinematical limit of the pair production process in $e^+e^-$ colliders. We investigate these mass ranges and discuss the possibility to test the renormalization group equations for the selectron masses.
\end{abstract}


\section{Introduction}

The electron-photon collision mode of an $e^+e^-$ linear collider \cite{PLC} 
provides us with the possibility to produce single selectrons in association 
with the lightest supersymmetric particle (LSP), which is assumed to be the 
lightest neutralino $\tilde{\chi}_1^0$. Thus selectrons can be produced with 
masses beyond the kinematical range for pair production at an $e^+e^-$ linear 
collider. Also the production mechanism (electron exchange in the s-channel
and selectron exchange in the t-channel) for associated selectron-neutralino
production is simpler than that for selectron pair production in $e^+e^-$ 
collisions. Assuming a common scalar mass $m_0$ at the unification point the
masses of the selectrons are related to the MSSM parameters $\tan\beta$ and 
$M_2$, the $SU(2)$ gaugino mass parameter, by renormalization group equations 
\cite{Polchinski}. In the MSSM quite generally the right selectron 
$\tilde{e}_R$ is lighter than the left selectron $\tilde{e}_L$. In extended 
SUSY models,
however, $\tilde{e}_R$ could be heavier than $\tilde{e}_L$ \cite{extsusy}.
 We study in this 
paper the process $e^-\gamma \longrightarrow \tilde{\chi}_1^0 
\tilde{e}_{L/R}^- \longrightarrow e^- \tilde{\chi}_1^0 \tilde{\chi}_1^0$ with 
polarized beams.
Since the cross section and the forward-backward asymmetry of the decay
electron depend sensitively on the selectron masses this process is
suitable for testing the renormalization group relation between 
$m_{\tilde{e}_R}$ and $m_{\tilde{e}_L}$.

\section{Cross Section and Forward-Backward Asymmetry}
The associated production of selectrons and the LSP proceeds
via $e^-$ exchange in the s-channel and $\tilde{e}_{L/R}$ exchange
in the t-channel.
In the narrow width approximation the total cross section 
$\sigma_{e\gamma}^{L/R}$ for the combined process of $\tilde{e}_{L/R}^-\tilde{\chi}_1^0$ production and the subsequent decay $\tilde{e}_{L/R}^-\longrightarrow e^-\tilde{\chi}_1^0$ factorizes into the production cross section $\sigma_P$
and the leptonic branching ratio:
\begin{equation}
\sigma_{e\gamma}^{L/R} = \sigma_P\left(s_{e\gamma}\right) \cdot BR\left(
\tilde{e}_{L/R}^- \longrightarrow e^-\tilde{\chi}_1^0\right)
\end{equation} 
The measured cross section $\sigma_{ee}^{L/R}$ in the $e^+e^-$ cms is obtained by folding 
$\sigma_{e\gamma}^{L/R}$ with the energy spectrum $P(y)$ of the Compton 
backscattered laser beam taking into account the mean helicity of the photon beam
\cite{Py}:
\begin{equation}
\sigma_{ee}^{L/R} = \int dy P\left(y\right)d\hat{\sigma}_{e\gamma}^{L/R}\left(s_{e\gamma}=ys_{ee}\right)
\end{equation}
\begin{equation} 
\hat{\sigma}_{e\gamma}^{L/R}=\sigma_{e\gamma}^{L/R}\left(1+\lambda\left(y\right) A_c\right)
\end{equation} 
$A_c$ is the circular photon asymmetry and $y=E_{\gamma}/E_e$ is the ratio of
the photon energy and the energy of the converted electron beam. The energy 
spectrum $P\left(y\right)$ and the mean helicity $\lambda\left(y\right)$ of 
the high energy photon beam sensitively depend on the polarizations $\lambda_L$
of the laser beam and $\lambda_k$ of the converted electron beam.
Beyond the cross section $\sigma_{ee}=\sigma_{ee}^L+
\sigma_{ee}^R$, 
we study the forward-backward asymmetry of the decay electrons:
\begin{equation}
A_{FB}=\frac{\sigma_{ee}^F-\sigma_{ee}^B}{\sigma_{ee}^F+\sigma_{ee}^B}
\end{equation}
The forward direction is
defined by the electron beam.

Apart from the kinematics the selectron masses enter the cross sections and 
the forward-backward asymmetries
explicitly via the selectron propagator in the t-channel. 
Assuming a common scalar mass $m_0$ at the unification point the            
masses of the selectrons are related to the MSSM parameters $\tan\beta$ and
the gaugino mass parameter $M_2$ by renormalization group equations
\cite{Polchinski}:
\begin{equation}
m^2_{\tilde{e}_R}=m_e^2+m_0^2+0.23M_2^2-m_Z^2\cos 2\beta\sin^2\theta_W
\end{equation}
\begin{equation}
m^2_{\tilde{e}_L}=m_e^2+m_0^2+0.79M_2^2+m_Z^2\cos 2\beta\left(-0.5+\sin^2\theta_W\right)
\end{equation}
 In the MSSM quite generally the right selectron                  
$\tilde{e}_R$ is lighter than the left selectron $\tilde{e}_L$. In extended SUSY
models these relations are changed  as a consequence of additional D-terms in the scalar
potential and the right selectron may be heavier than the left selectron.

In chapter 3 we study the dependence of the cross section $\sigma_{ee}$ and the 
forward-backward asymmetry $A_{FB}$ on $m_{\tilde{e}_R}$ and $m_{\tilde{e}_L}$. We 
shall see that this process is useful for testing the GUT-relations equs. (5), (6).

\section{Numerical Results}

We present numerical results for the MSSM parameters $M_2=152$ GeV, $M_1=78.7$ 
GeV, $\mu=316$ GeV and $\tan\beta=3$ for the cms-energy $\sqrt{s_{ee}}=500$ 
GeV. The LSP is gaugino-like  with $m_{\tilde{\chi}_1^0}=71.9$ GeV. For 
$m_{\tilde{e}_R}=127$ GeV and $m_{\tilde{e}_L}=171$ GeV this corresponds to 
one ECFA/DESY reference scenario for the linear collider 
\cite{refscen1}. Fig. 1 shows the total cross section 
$\sigma_{ee}$ and the forward-backward asymmetry $A_{FB}$ for $\lambda_k=+1$ 
and $\lambda_L=-1$.
This choice of $\lambda_k$ and $\lambda_L$ leads to a strongly marked  high 
energetic 
peak in the energy spectrum $P(y)$ \cite{Py} and therefore to maximal cross 
sections. For the electron beam in the $e\gamma$ collision we  choose in fig. 1a,b
the polarization $P_e=0.9$. Then due to $\sigma_{ee}^{L/R}\propto 
\left(1\mp P_e\right)$ the cross section for $\tilde{e}_R$ is enhanced whereas
that for $\tilde{e}_L$ is strongly suppressed so that $\sigma_{ee}$ is nearly 
independent of $m_{\tilde{e}_L}$. The cross section for this polarization
configuration (fig. 1a) allows to constrain $m_{\tilde{e}_R}$ up to 344 GeV. 
In a region around 200 GeV the dependence of $\sigma_{ee}$ on $m_{\tilde{e}_R}$
is rather weak. In this case $A_{FB}$ (fig. 1b)  gives 
additional informations on the mass $m_{\tilde{e}_R}$. For 
$m_{\tilde{e}_R}>344$ GeV the production of right selectrons becomes 
impossible and due to the suppression by the polarization factor 
$\left(1-P_e\right)$ the cross section is rather small: $\sigma_{ee}=\sigma_{ee}^L\sim 2.5$ fb for $m_{\tilde{e}_L}=100$ GeV. 
Then $A_{FB}$ only depends on $\sigma_{ee}^L$ and is 
independent of $m_{\tilde{e}_R}$ (see fig. 1b).

For figs. 1c and 1d we choose $P_e=-0.9$. Now the production and
decay of left selectrons is no longer neglectible. Fig. 1c gives
the cross sections for three different masses $m_{\tilde{e}_R}=100$ GeV, 127 GeV, 200 GeV. In all three cases it should be possible to constrain 
$m_{\tilde{e}_L}$ up to 170 GeV. For masses larger than 170 GeV the
dependence of $\sigma_{ee}$ on $m_{\tilde{e}_L}$ is too weak so that one obtains
only a lower limit on $m_{\tilde{e}_L}$.
For large values of $m_{\tilde{e}_R}$ the measurement of the asymmetry 
$A_{FB}$ (fig. 1d) could be helpful to extend this mass range to somewhat 
higher values.

\setlength{\unitlength}{1cm}
\begin{figure}[t]
\label{f1LRhoch4}
\centering
\begin{picture}(12.6,9.2)
\put(-0.9,2.2){\includegraphics{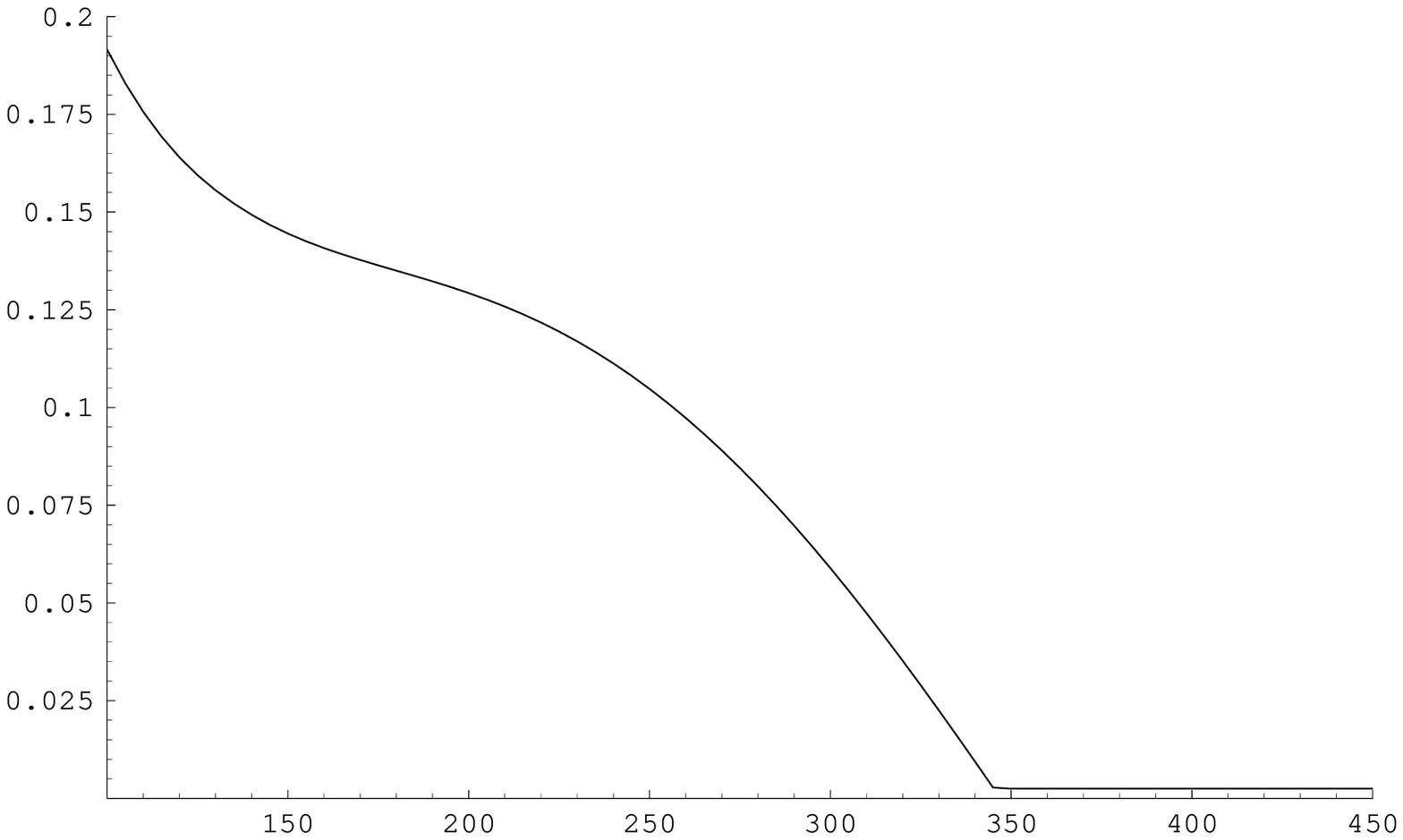}}
\put(5.2,5.2){{\tiny $m_{\tilde{e}_R}$/GeV}}
\put(0.0,9.0){{\tiny $\sigma_{ee}$/pb}}
\put(2.5,5.1){{\tiny (a)}}
\put(5.5,2.2){\includegraphics{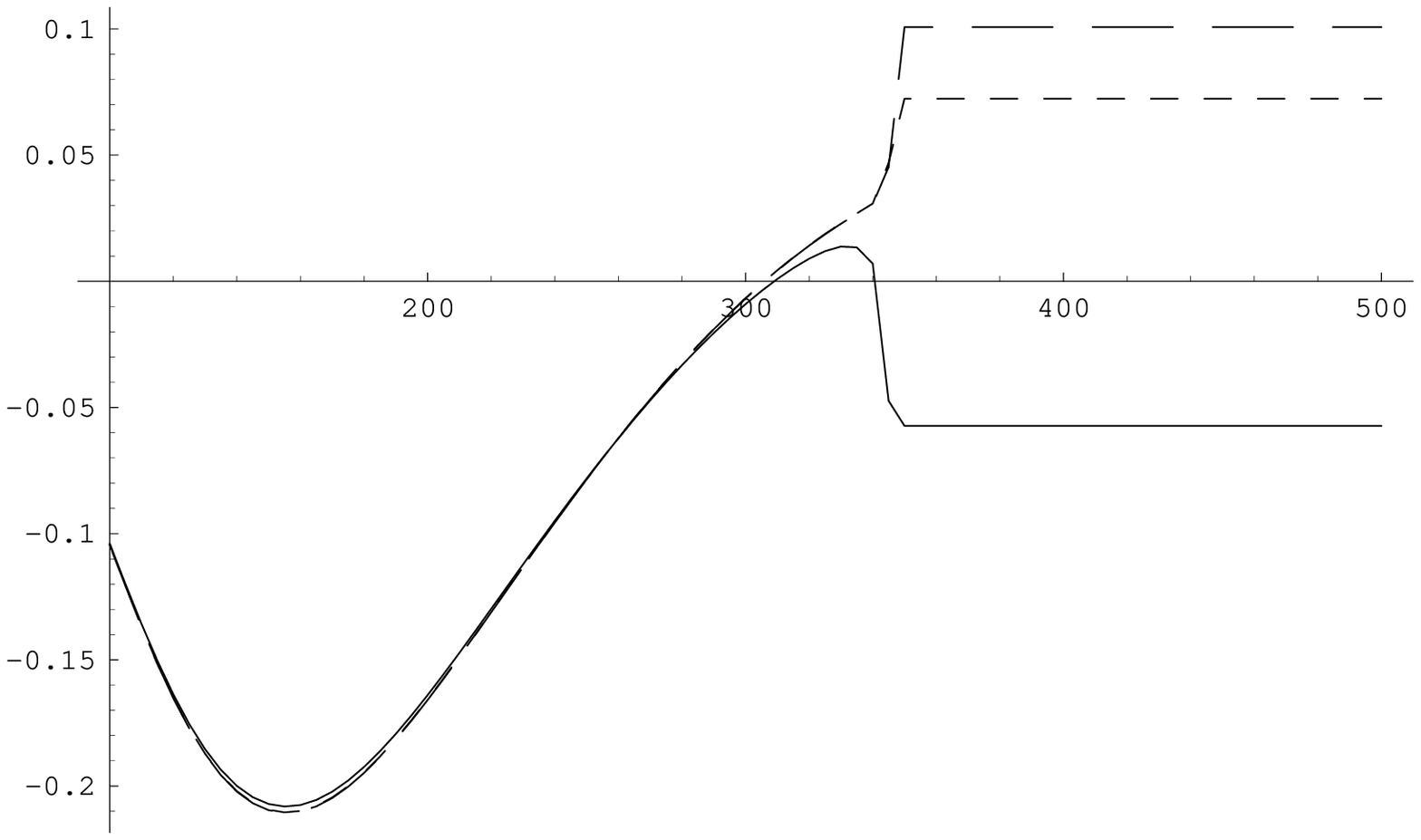}}
\put(11.4,7.9){{\tiny $m_{\tilde{e}_R}$/GeV}}
\put(6.5,9.0){{\tiny $A_{FB}$}}
\put(7.19,5.8){\circle{0.1}}
\put(9.2,5.1){{\tiny (b)}}
\put(-0.9,-2.6){\includegraphics{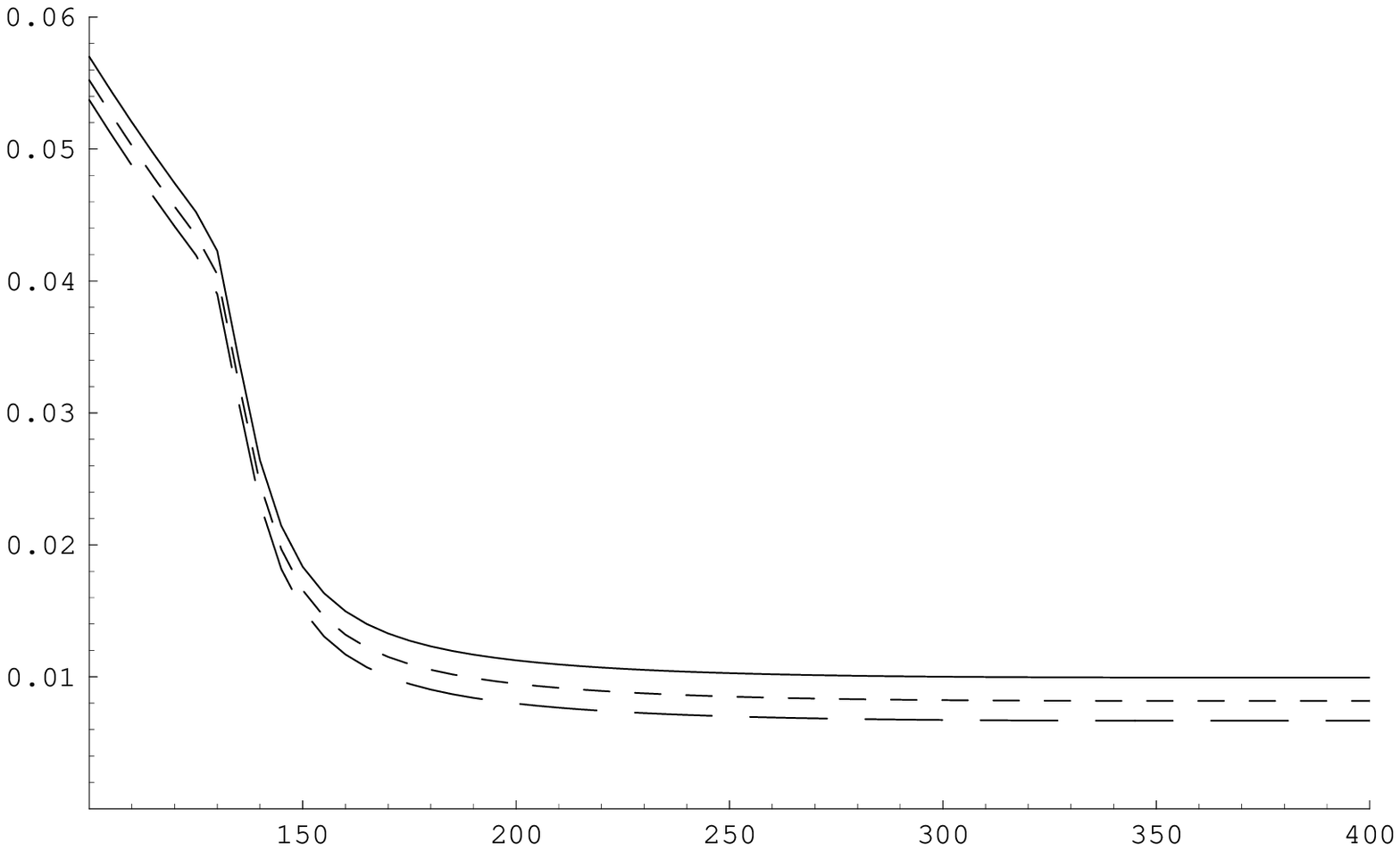}}
\put(5.2,0.3){{\tiny $m_{\tilde{e}_L}$/GeV}}
\put(0.0,4.2){{\tiny $\sigma_{ee}$/pb}}
\put(1.62,1.33){\circle{0.1}}
\put(2.3,0.1){{\tiny (c)}}
\put(5.5,-2.6){\includegraphics{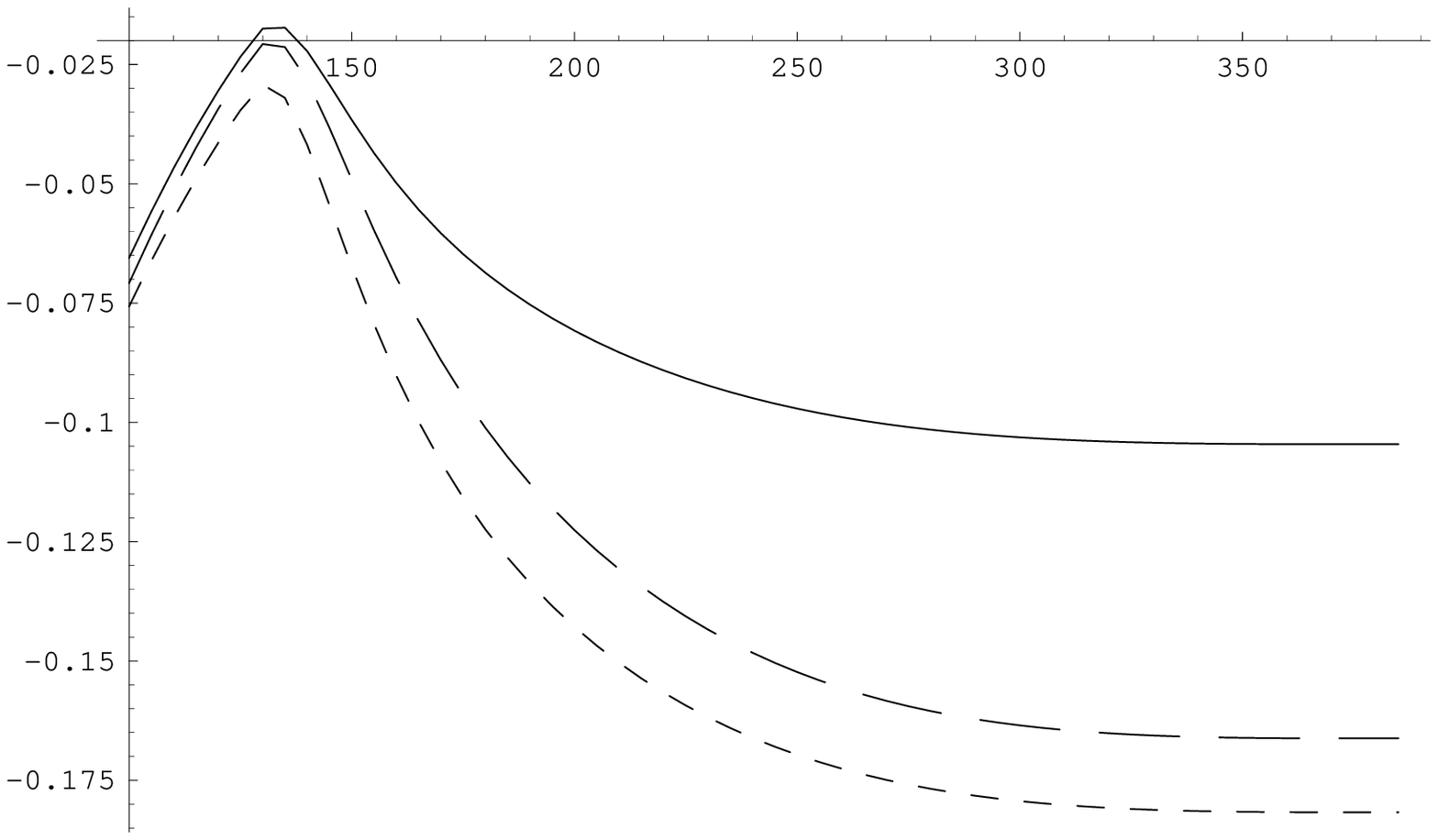}}
\put(11.4,4.1){{\tiny $m_{\tilde{e}_L}$/GeV}}
\put(6.5,4.2){{\tiny $A_{FB}$}}
\put(8.19,2.12){\circle{0.1}}
\put(9.2,0.1){{\tiny (d)}}
\end{picture}
\caption{Cross sections and forward-backward asymmetries for
$\sqrt{s}_{ee}=500$ GeV, $\lambda_k=+1$, $\lambda_L=-1$ (the values of the
ECFA/DESY reference scenario are marked by small circles); (a) dependence
of the total cross section $\sigma_{ee}$ on $m_{\tilde{e}_R}$ for
$P_e=0.9$ and $m_{\tilde{e}_L}=100$ GeV (nearly independetnt of $m_{\tilde{e}_L}$); (b) dependence of the asymmetry
$A_{FB}$ on $m_{\tilde{e}_R}$ for $P_e=0.9$ ($m_{\tilde{e}_L}=100$ GeV
--------, $m_{\tilde{e}_L}=171$ GeV - - - -, $m_{\tilde{e}_L}=200$ GeV ---
--- ---); (c) dependence of the total cross section $\sigma_{ee}$ on
$m_{\tilde{e}_L}$ for $P_e=-0.9$ ($m_{\tilde{e}_R}=100$ GeV --------, $m_{\tilde{e}_R}=127$ GeV - - - -, $m_{\tilde{e}_R}=200$ GeV --- --- ---); (d) dependence of the asymmetry $A_{FB}$ on $m_{\tilde{e}_L}$ for $P_e=-0.9$ ($m_{\tilde{e}_R}=100$ GeV --------, $m_{\tilde{e}_R}=127\mbox{ GeV}  - - - -$, $m_{\tilde{e}_R}=200$ GeV --- --- ---)}
\end{figure}

If the renormalization group relations equs. (5), (6) are satisfied then $m_{\tilde{e}_L}$ is larger
than $m_{\tilde{e}_R}$, $m_{\tilde{e}_L}^2-m_{\tilde{e}_R}^2 \sim  0.56 M_2^2$.
This relation can be tested with the total cross sections in figure 1a and 
1c up to  $m_{\tilde{e}_L}=170$ GeV, complementary to the
measurements at an $e^+e^-$-collider. Fig. 1d shows that for higher masses of
$\tilde{e}_R$ the asymmetry could allow a test of the renormalization group
relations, equs. (5),(6).

\section{Conclusion}
With a suitable choice of beam polarizations it is possible to constrain
$m_{\tilde{e}_R}$ up to 344 GeV and $m_{\tilde{e}_L}$ up to 170 GeV in the
process $e^-\gamma \longrightarrow \tilde{\chi}_1^0 \tilde{e}_{L/R}^- 
\longrightarrow e^- \tilde{\chi}_1^0 \tilde{\chi}_1^0$ from a measurement of
the total cross 
sections. The forward-backward asymmetry $A_{FB}$ gives additional information
on these masses. Especially one could measure masses $m_{\tilde{e}_L}>170$
GeV
if $m_{\tilde{e}_R}$ is high enough. The cross sections and the 
forward-backward asymmetries allow to test the equations
for the selectron masses 
complementary to an $e^+e^-$-collider.

\section{Acknowledgements}

We are grateful to Gudrid Moortgat-Pick and Stefan Hesselbach for valuable
discussions. Claus Bl\"ochinger thanks the organizers of the XXIII. School
of Theoretical Physics for the friendly atmosphere during the conference.
This work was supported by the Deutsche Forschungsgemeinschaft under contract no. FR 1064/4-1, the Bundesministerium f\"ur Bildung und Forschung (BMBF) under contract number 05 HT9WWA 9 and the Stiftung f\"ur deutsch-polnisch Zusammenarbeit, Warszawa.

\end{document}